\documentclass{ws-ijmpa}
\usepackage[super,compress]{cite}
\usepackage{graphicx}
\begin{document}

%
\catchline{}{}{}{}{}
%

\title{VISCOSITIES OF GLUON DOMINATED QGP MODEL WITHIN RELATIVISTIC NON-ABELIAN HYDRODYNAMICS}

\author{T.P. DJUN}
\address{Graduate Programs in Material Science, 
  University of Indonesia, 
  Kampus UI Salemba, \\
 Jakarta 10430, Indonesia \\
tpdjun@teori.fisika.lipi.go.id}

\author{L.T. HANDOKO}
\address{
Group for Theoretical and Computational Physics,
  Research Center for Physics, Indonesian Institute of Sciences,
  Kompleks Puspiptek Serpong, \\
  Tangerang 15310, Indonesia}

\author{B. SOEGIJONO}
\address{
Graduate Programs in Material Science, 
  University of Indonesia, 
  Kampus UI Salemba, \\
 Jakarta 10430, Indonesia }

\author{T. MART}
\address{
Departemen Fisika, FMIPA, Universitas 
  Indonesia, \\
  Depok 16424, Indonesia}
\maketitle


\begin{abstract}
Based on the first principle calculation, a Lagrangian for the 
system describing quarks, gluons, and their interactions, is constructed. 
Ascribed to the existence of dissipative behavior as a 
consequence of strong interaction within quark-gluon plasma (QGP) matter, 
auxiliary terms describing viscosities are constituted 
into the Lagrangian. Through a "kind" of phase transition, 
gluon field is redefined as a scalar field with four-vector 
velocity inherently attached. Then, the Lagrangian is 
elaborated further to produce the energy-momentum tensor 
of dissipative fluid-like system and the equation 
of motion (EOM). By imposing the law of 
energy and momentum conservation, the values of shear and 
bulk viscosities are analytically calculated. Our result shows that, at the energy level 
close to hadronization,  the bulk viscosity is bigger than shear viscosity. 
By making use of the conjectured values  $\eta / s \sim 1 / 4\pi$ and $\zeta / s \sim 1$,  
the ratio of bulk to shear viscosity is found to be $\zeta / \eta > 4 \pi$.

\end{abstract}

\ccode{PACS numbers: 66.20.+d, 24.85.+p, 12.38.Mh}


\section{Introduction}	

Relativistic heavy ion collision experiments have confirmed that a strongly coupled quark-gluon plasma (QGP) has been reproduced. Lots of new phenomena from this new state of matter, like its small transport coefficients and the real phase transition mechanism from deconfined quarks to confined hadron, have currently become hot research topics. 

Strong interaction typically indicates the existence of small 
transport coefficients of a system in order to ensure the 
efficiency of momentum transmission. In such kind of system, 
shear and bulk viscosities are naturally become its 
inherent properties. Understanding these properties firmly 
has been considered  
as a prerequisite knowledge for dynamical description of plasmas, 
astrophysics, early universe cosmology, and other branches of 
physics.  

The dynamics of quark-gluon plasma that intuitively been considered 
to live in the realm of quantum chromodynamics (QCD) has been studied through the lattice 
theory \cite{Gottlieb:2007zz,Petreczky:2007bn}. Nevertheless, some 
simplifications are always required in order to reduce 
the calculation complexities as well as to cope with the 
limited computing capacity of the current state-of-the-art computer.
On the other hand, experiments of relativistic heavy-ion collision 
also exhibit a collective flow behavior. The typical character 
which belongs to fluid dynamics is considered as one of the validating 
aspect for QGP to be modeled as a relativistic 
hydrodynamics system \cite{Bouras:2009nn,Romatschke:2009im,Teaney:2000cw,
Huovinen:2001cy,Kolb:2001qz,Kolb:2002ve,Hirano:2002ds,Baier:2006gy}. 
In the development, the discovery of 
new phenomena from experiment such as the shock wave in form of Mach 
cone has guided researchers to materialize the gluonic-plasma model 
\cite{Bluhm:2010qf,Adams:2003kv,Adare:2008qa}.  
Recent achievements in hydrodynamics description of 
relativistic heavy-ion collision have inspired more researchers to 
use relativistic fluid dynamics approach on searching for viscosities 
of QGP. 

There is also a hybrid model that describes the relativistic hot plasma by unifying 
electromagnetic fields with flow field. The unification is represented 
by the effective field strength tensor, 
$M_{\mu\nu} = F_{\mu\nu} + (m/q) S_{\mu\nu}$, combining appropriately 
the electromagnetic and fluid fields \cite{Bambah:2006yg}.  

In this paper, we follow this hybrid concept, except that all 
fields are constructed by following the first principle. 
In this framework, the strongly interacting system of QCD is considered as a 
macroscopic fluid system rather than the result of many-particle interaction. 
This scenario is, somehow, inspired by the works of Jackiw {\it et al}.
\cite{jackiw,jackiw2}, where they took an analogy of a non-commutative 
quantum system in developing a mathematical description of fluid structure, 
that is still consistent when applied to the microscopic fluid system 
(non-Abelian fluid).  Instead of  taking the same route, we start with different 
way, i.e., by following the first principle. To this end, a viscous fluid-like QCD 
Lagrangian is constructed. By redefining the gluon fields, a sort of phase 
transformation on the Lagrangian appears. We then obtain an expression that 
can be interpreted as a microscopic interacting fluidic system. 
This strategy also produces a hybridization that combines the flow field and 
the charge field in one system. Further, by making use of
the Lagrangian, the formulation for bulk and shear viscosities can be obtained. 
At the energy level near to the phase transition, the result of our calculation 
exhibits a large ratio of bulk to shear viscosity, just as what is being seen 
at RHIC. This result corroborates the new findings on QCD plasma.

Our paper is organized as follows. In Sect.~\ref{sec:background} 
we briefly review the basic theory of fluid QCD, and explain the 
phase transition from the deconfined particle field to the flow 
field. In Sect.~\ref{sect:formula}, the shear and bulk viscosities 
of gluon dominated QGP are formulated as a function of the energy 
density. The obtained results will be discussed in 
Sect.~\ref{sect:result}. Finally, we will summarize our finding
in Sect.~\ref{sect:summary}. Throughout this work we use 
the natural units, i.e., $\hbar \equiv c \equiv \; 1$.

\section{Theoretical background: The gluon dominated QGP}
\label{sec:background}

In general, the Lagrangian of the viscous QGP can be assumed to consist of a 
nondissipative (ideal) and a dissipative part. Mathematically, we can
write 
\begin{eqnarray}
  \label{eq:definition}
  {\cal L} = {\cal L}_{\rm ideal} + {\cal L}_{\rm diss.} ~,
\end{eqnarray}
where the ideal part reads~\cite{sulaiman},
\begin{equation}
{\cal L}_{\rm ideal} = i \bar{Q} \gamma^\mu \partial_\mu  Q - m_Q \bar{Q} Q - 
{\textstyle \frac{1}{4}} S^a_{\mu\nu} S^{a\mu\nu}  - {\textstyle \frac{1}{4}} F_{\mu\nu} F^{\mu\nu} + g_{F} 
J^a_{F \mu} U^{a\mu} + g_G J_{G \mu} A^{\mu} .
\label{eq:Lagrangian-ideal}
\end{equation}
Here $Q$ and $\bar{Q}$ represent the quark and anti-quark triplet, 
$\gamma^\mu$ is Dirac matrices, and $m_Q$ is the quark mass. 
The kinetic term for gauge boson $A_\mu$ is built inside its field 
strength tensor, i.e.,
$F_{\mu\nu} \equiv  \partial_\mu A_\nu - \partial_\nu A_\mu $. 
The same role is taken for the gauge boson $U^a_\mu$ of SU(3) gauge 
group,  the kinetic terms $S^a_{\mu\nu} \equiv \partial_\mu U^a_\nu - 
\partial_\nu U^a_\mu + g_F f^{abc} U^b_\mu U^c_\nu$, where $g_F$ is the strong 
coupling constant. The last two terms, $J^{a}_{F \mu} = \bar{Q} T^a_F 
\gamma_\mu Q$ and $J_{G \mu} = \bar{Q} \gamma_\mu Q$, represent the quark 
currents from the SU(3) and $U$(1) gauge groups, respectively, and $g_G$ denotes 
the coupling constant from $U$(1) gauge group.

Being a Lagrangian of QCD, ${\cal L}_{\rm ideal}$ is constructed from 
the fact that the derived energy-momentum tensor 
has an identical form as the energy-momentum tensor of an 
ideal fluid, as shown by Ref. 18.
It is the dominant part of the Lagrangian 
that describes the unification of fermions and bosons from different gauge 
groups, that preserves the ${\rm SU}(3)_{F} \otimes {\rm U}(1)_{G}$ 
gauge symmetry. It represents a non-Abelian SU(3) fluid fields containing quarks and 
anti-quarks that interact with an electromagnetic field. 

The nonideal or dissipative part of Eq.~(\ref{eq:definition}) can be
written as,
\begin{eqnarray}
{\cal L}_{\rm diss.} &=& - \eta T^a ( \partial^\mu U^a_\mu +    \partial^\nu U^a_\nu -  U^{a \mu} U^{a \varepsilon} \partial_\varepsilon U^a_\mu -   U^{a \nu} U^{a \varepsilon} \partial_\varepsilon U_{a \nu}   ) \nonumber\\
&&-  T^a (\zeta - \frac{2}{3} \eta) \partial_\varepsilon U^{a \varepsilon} ( 1 - U^{a \nu} U^a_\nu  ) ~. 
\label{eq:Lagrangian-diss}
\end{eqnarray}
This part is  constructed from the relativistic hydrodynamics viscous 
energy-momentum tensor developed by Landau and others \cite{landau, weinberg}. 
In Eq.~(\ref{eq:Lagrangian-diss}), 
$\eta$ and $\zeta$ denote the shear and bulk viscosities, respectively,
whereas $T^a$ is the structure constant of the SU(3) gauge group.\\
Therefore, the complete Lagrangian can be written as
\begin{eqnarray}
{\cal L} &=& i \bar{Q} \gamma^\mu \partial_\mu  Q - m_Q \bar{Q} Q - 
{\textstyle \frac{1}{4}} S^a_{\mu\nu} S^{a\mu\nu}  - {\textstyle \frac{1}{4}} F_{\mu\nu} F^{\mu\nu} + g_{F} 
J^a_{F \mu} U^{a\mu} + g_G J_{G \mu} A^{\mu} \nonumber\\
&& - \eta T^a ( \partial^\mu U^a_\mu +    \partial^\nu U^a_\nu -  U^{a \mu} U^{a \varepsilon} \partial_\varepsilon U^a_\mu -   U^{a \nu} U^{a \varepsilon} \partial_\varepsilon U_{a \nu}   ) \nonumber\\
&&-  T^a (\zeta - \frac{2}{3} \eta) \partial_\varepsilon U^{a \varepsilon} ( 1 - U^{a \nu} U^a_\nu  ) ~.
\label{eq:Lagrangian}
\end{eqnarray}

By introducing the viscous terms into the Lagrangian, it is corollary 
representing a "near to ideal" fluid, while the system itself is near to the 
equilibrium.  Gauge symmetry of the Lagrangian is preserved until 
the dissipative terms are added. 
This is understandable as the fact that gauge invariance is broken 
when the fluid-like system leaves its ideal condition. \\
\indent To introduce the flow property into the fluid field, a construction of 
relativistic form is in order. It has been argued that 
$U^a_{\mu}$ should be rewritten in  a particular form of the relativistic 
velocity as $U^a_\mu = (U^a_0 , \textbf{U}^a) \equiv u_\mu \phi^a$ 
\cite{sulaiman,djun}. Here, $u_\mu \equiv \gamma  (1, 
\textbf{v} )$ and $\gamma  = (1-\vert\textbf{v} \vert^2)^{-1/2}$, 
with $\phi^a = \phi^a (x)$ represents a dimension one scalar field. 
Equation of motion (EOM) for a single gluon field from the above Lagrangian 
leads to a general relativistic fluid equation \cite{sulaiman}. 
This fact explains that a single gluonic field $U^a_\mu$ may behave 
as a fluid at certain phase and as a point particle at hadronic 
state with a polarization vector $\epsilon_\mu$ in the conventional 
form of $U^a_\mu = \epsilon_\mu \phi^a$. This is considered as a 
kind of ''phase transition'' between fluidic and hadronic states. 
\begin{equation*}
   \underbrace{\rm hadronic~ state}_{\epsilon_\mu }  \longleftrightarrow 
   \underbrace{\rm QGP~ state}_{u_\mu } ~.
\end{equation*}
A more detailed discussion on this topic can be found in Ref. 21.

By applying the Euler-Lagrange equation to Eq.~(\ref{eq:Lagrangian}), and 
ignoring the minor contribution from the dissipation part, we obtain 
\begin{equation}
\frac{\partial}{\partial t} (\gamma \mathbf{v} \phi^a) + \nabla 
(\gamma \phi^a) = -g_F \oint d\mathbf{x} (J^a_{F 0} + F^a_0).
\label{eq:eq-of-motion}
\end{equation}
In the non-relativistic limit, i.e., $\gamma \sim 1 
+ \frac{1}{2} \vert \mathbf{v} \vert^2 $ and $\phi^a \sim 1$, for 
a constant $\phi^a$, Eq.~(\ref{eq:eq-of-motion}) reduces to 
the classical equation of motion 
of fluid dynamics, i.e.,
\begin{equation}
\frac{\partial \mathbf{v} }{\partial t} + (\mathbf{v} \cdot \nabla) 
\mathbf{v} = -g_F \oint d\mathbf{x} (J^a_{F 0} + F^a_0)\mid_{\rm non-rel} ,
\label{eq:eq-of-motion-nonrel}
\end{equation}
where 
$J^a_{F 0}$ is a covariant current induced by the existing 
quarks surrounded by and interacting with gluon "fluid", while 
$F^a_\mu$ is induced by the fluid self-interaction and the interacting 
gauge fields $A^a_\mu$. 
This shows that from a certain point of view the Lagrangian describes 
a general relativistic fluid system interacting with another gauge 
field and matter inside.  Besides that it also fulfills the 
basic presupposed idea that in comparison with the size of a system, 
the relatively long-distance, low frequency behavior of any interacting 
theory at finite temperature could be described by the  theory of fluid 
mechanics \cite{landau}. 

From Eq.~(\ref{eq:eq-of-motion-nonrel}) it is obvious that the flow 
characteristic describing
a plasma is represented by the gluon fields and, accordingly, the system 
that we are working with is a gluon dominated QGP. Therefore, 
all nongluon terms or terms that have no 
interaction with gluon can be omitted.  As a consequence, 
${\cal L}_{\rm diss}$ does not change
and  ${\cal L}_{\rm ideal}$ is simplified. Now the Lagrangian reads 
\begin{eqnarray}
{\cal L} &=& - {\textstyle \frac{1}{4}} S^a_{\mu\nu} S^{a\mu\nu}  + g_F J^a_{F \mu} U^{a \mu} \nonumber\\
&& - \eta T^a ( \partial^\mu U^a_\mu +    \partial^\nu U^a_\nu -  U^{a \mu} U^{a \varepsilon} \partial_\varepsilon U^a_\mu -   U^{a \nu} U^{a \varepsilon} \partial_\varepsilon U_{a \nu}   ) \nonumber\\
&&-  T^a (\zeta - \frac{2}{3} \eta) \partial_\varepsilon U^{a \varepsilon} ( 1 - U^{a \nu} U^a_\nu  ) ~.
\label{eq:qgp-Lagrangian}
\end{eqnarray}
Here ${\cal L}_{\rm ideal} = - {\textstyle \frac{1}{4}} S^a_{\mu\nu} S^{a\mu\nu}  + g_F J^a_{F \mu} U^{a \mu}$.
Besides describing the kinetics of gluons, it also 
indicates the self-interaction between small number of quark and anti-quark, the 
self-interaction between gluons, and the interaction between quark with 
gluon ''fluid'', where it is immersed in.  It should be noted that the quarks 
and anti-quarks feel the electromagnetic force due to the $U(1)$ field $A_\mu$, 
but the size is suppressed by a factor of $e/g = \sqrt{\alpha / \alpha_s} \backsim O(10^{-1})$.

\section{Shear and bulk viscosities}
\label{sect:formula}

From the Lagrangian one can also derive the energy momentum tensor 
density by means of ${\cal T}_{\mu\nu} = \frac{2}{\sqrt{-g}} \frac{\delta ( {\cal L} \sqrt{- g})}{\delta 
g^{\mu\nu}} $. For the ideal term ${\cal L}_{\rm ideal}$, 
one immediately gets \cite{nugroho}
\begin{eqnarray}
\mathcal{T}_{\mu\nu \;\; ideal} &=& S^a_{\mu\rho} S^{a \rho}_{\;\;\;\; \nu} - g_{\mu\nu} {\cal L} + 2 g_F J^a_{F \mu} U^a_\mu \nonumber\\
&=& [  2 g_F g_{\mu\nu} J^a_{F \mu} U^{a \mu} +  g^2_F f^{abc} f^{ade} U^b_\mu U^c_\rho U^{d \rho} U^e_\nu ] \nonumber \\
&& - [ g_F g_{\mu\nu}  J^a_{F \mu} U^{a\mu} -  \frac{1}{4} g_{\mu\nu} g^2_F f^{abc} f^{ade} U^b_\mu U^c_\nu U^{d \mu} U^{e \nu} ] ~.
\label{eq:energy-momentum-tensor}
\end{eqnarray}

\noindent $J^a_{F \mu} U^{a\mu}$ in Eq.(\ref{eq:energy-momentum-tensor}) can be obtained by considering the EOM (Dirac equation) of a single colored quark $(Q)$ or anti-quark $(\bar{Q})$ with 4-momentum $p_\mu$. Since the solution of the EOM is $Q (p, x) = q (p)$ exp$(-ip \cdot x)$, one immediately gets $\bar{q} \gamma_\mu q = 4 p_\mu$. Assuming that all colored quarks/anti-quarks have the same momenta and the velocity of gluons and quarks inside the gluon sea are homogeneous, approximately $J^a_{F \mu} U^{a \mu} \propto 4 p_\mu T^a U^\mu = 4 m_Q T^a \phi^a $ since $u_\mu u^\mu = u^2 = 1$. Then the ideal part of energy momentum tensor in the function of field $\phi^a$ reads,
\begin{eqnarray}
\mathcal{T}_{\mu\nu \;\; ideal} &=& [ 8 g_F m_Q T^a \phi^a + g^2_F f^{abc} f^{ade} \phi^b \phi^c \phi^d \phi^e  ] u_\mu u_\nu \nonumber \\
&&- [ 4 g_F m_Q T^a \phi^a  -  \frac{1}{4} g^2_F f^{abc} f^{ade} \phi^b \phi^c \phi^d \phi^e ] g_{\mu\nu} ~.
\label{eq:energy-momentum-tensor_ideal}
\end{eqnarray}
After obtaining  ${\cal T}_{\mu\nu \;\; ideal}$  that represents a perfect fluid, 
we can derive the dissipative part of energy-momentum tensor ${\cal T}_{\mu\nu \;\; diss}$ from
 ${\cal L}_{diss}$, i.e.,
\begin{eqnarray}
{\cal T}_{\mu\nu \;\; diss} &=& - \eta T^a (\partial_\nu U^a_\mu + \partial_\mu U^a_\nu - 
  U^a_\nu U^{a \varepsilon} \partial_\varepsilon U^a_\mu - U^a_\mu U^{a \varepsilon} \partial_\varepsilon U^a_\nu) \nonumber\\
&&+ {\textstyle \frac{2}{3}} 
  \eta T^a  \partial_\varepsilon U^{a \varepsilon} (g_{\mu\nu} - U^a_\mu U^a_\nu) 
- \zeta T^a \partial_\varepsilon U^{a \varepsilon} (g_{\mu\nu} - U^a_\mu U^a_\nu) .
\end{eqnarray} 
The structure of the outcomes reveals a typical form of the relativistic 
hydrodynamics viscous energy-momentum tensor 
\cite{landau, weinberg}. 
Further, we assume that the gluon color states are 
homogeneous, i.e., $\phi^a = \phi$ for all $a = 1,\cdots,8$.
As the follow-on to this assumption, 
the generator $T^a$ can be compactly written as $T = \sum_{a} T^a$,
and also the 2nd and 4th 
terms of $\mathcal{T}_{\mu\nu \;\; ideal}$ can be omitted due to the completely 
anti-commutative property of the structure constant $f^{abc}$.
Now, the complete energy momentum tensor that consist of ideal and 
dissipative part  reads,
\begin{eqnarray}
{\cal T}_{\mu\nu} &=& \mathcal{T}_{\mu\nu \;\; ideal} + \mathcal{T}_{\mu\nu \;\; diss} \nonumber\\
 &=& (8 T g_F m_Q \phi) u_\mu u_\nu -  (4 T g_F m_Q \phi) g_{\mu\nu} \nonumber\\
&& -  \eta T \Big( \partial_\nu (u_\mu \phi) + \partial_\mu (u_\nu \phi) - (u_\nu \phi) (u^\varepsilon \phi) \partial_\varepsilon (u_\mu \phi) - (u_\mu \phi) (u^\varepsilon \phi) \partial_\varepsilon (u_\nu \phi) \Big) \nonumber\\
&& + \frac{2}{3}  \eta T \partial_\varepsilon (u^{ \varepsilon} \phi)  g_{\mu\nu} (1 - \phi^2 )  -   \zeta T \partial_\varepsilon (u^{ \varepsilon} \phi)  g_{\mu\nu} (1 - \phi^2 ) 
\label{eq:energymomentumtensor}
 \end{eqnarray}
The total energy momentum tensor is obtained by integrating 
Eq.~(\ref{eq:energymomentumtensor}) over the total volume 
of the space-time under 
consideration. It reveals the collective gluons flow 
in the system. 

The shear and bulk viscosities can be obtained
by imposing the energy and momentum conservation to the 
energy-momentum tensor. The conservation of energy is expressed 
as $\partial_0 {\cal T}^{00} + \partial_k {\cal T}^{0k} = 0$ \cite{ryder}. 
After some algebra and assuming that $\partial_\mu u^\mu = 0$ at 
a very short traveling distance and very short existence 
time of QGP, the conservation of energy can 
be formulated as 
\begin{equation}
\eta  a^0 + \zeta b^0 = c^0 , 
\label{eq:conserve-energy}
\end{equation}
where
\begin{eqnarray*}
c^0 &=& 8  g_F m_Q \Big[ (\partial_0 \phi) u^0 u^0  -  \frac{1}{2} (\partial_0 \phi) g^{00} + (\partial_k \phi) u^0 u^k)  \Big] ,\\
\\
a^0 &=& 2 \partial_0 \partial^0 (u^0 \phi) - 2 \big( u^0 u^\varepsilon (\partial_0 \phi^2) (\partial_\varepsilon \phi) + u^0 u^0 u^\varepsilon \phi^2 ( \partial_0 \partial_\varepsilon  \phi)  \big) \nonumber\\
&&- \frac{2}{3} \big( \partial_0 \partial_\varepsilon (u^{ \varepsilon} \phi) \big)  g^{00} (1 - \phi^2 )  + \frac{2}{3} \big( \partial_\varepsilon (u^{ \alpha} \phi) \big) g^{00} (\partial_0 \phi^2) + \partial_k \partial^k (u^0 \phi) \nonumber\\
&&+ \partial_k \partial^0 (u^k \phi)  - u^k  u^\varepsilon (\partial_k \phi^2) \partial_\varepsilon (u^0 \phi)  - u^k  u^\varepsilon \phi^2 \big( \partial_k \partial_\varepsilon (u^0 \phi) \big) \nonumber\\
&& - u^0 u^\varepsilon (\partial_k  \phi^2) \big( \partial_\varepsilon (u^k \phi) \big) - u^0 u^\varepsilon \phi^2 \big( \partial_k \partial_\varepsilon (u^k \phi) \big), \\
\\
b^0 &=& \big( \partial_0 \partial_\varepsilon (u^{ \varepsilon} \phi) \big) g^{00} (1 - \phi^2 ) -  \big( \partial_\varepsilon (u^{ \varepsilon} \phi) \big) g^{00} \big( \partial_0 \phi^2 \big). \\
\end{eqnarray*}
The conservation of momentum, 
$\partial_0 {\cal T}^{i0} + \partial_i {\cal T}^{ii} + \partial_k {\cal T}^{ik} = 0$, 
can be re-expressed in a simpler form as
\begin{equation}
\eta  x^i +  \zeta y^i = z^i ,
\label{eq:conserve-momentum}
\end{equation}
where
\begin{eqnarray*}
z^i &=& 8  g_F m_Q \big[ ( \partial_0 \phi) u^i u^0  +  (\partial_i \phi) u^i u^i - \frac{1}{2} (\partial_i \phi) g^{ii} +  ( \partial_k \phi) u^i u^k  \big],
\end{eqnarray*}
\begin{eqnarray*}
x^i &=&  \partial_0 \partial^0 (u^i \phi) + \partial_0 \partial^i (u^0 \phi) \nonumber\\
&&- u^0 u^\varepsilon (\partial_0 \phi^2) \partial_\varepsilon (u^i \phi) - u^0 u^\varepsilon \phi^2 \partial_0 \partial_\varepsilon (u^i \phi)  \nonumber\\
&&- u^i u^\varepsilon (\partial_0 \phi^2)  \partial_\varepsilon (u^0 \phi) - u^i u^\varepsilon \phi^2  \partial_0 \partial_\varepsilon (u^0 \phi)  \nonumber\\
&&+  2 \partial_i \partial^i (u^i \phi)  - 2 u^i u^\varepsilon (\partial_i \phi^2) \partial_\varepsilon (u^i \phi) - 2 u^i u^\varepsilon \phi^2 \partial_i \partial_\varepsilon (u^i \phi) \\
&& - \frac{2}{3} \big( \partial_i \partial_\varepsilon (u^{ \varepsilon} \phi) \big) g^{i i} (1 - \phi^2 ) + \frac{2}{3} \big(  \partial_\varepsilon (u^{ \varepsilon} \phi) \big)  g^{i i} (\partial_i \phi^2 )  +   \partial_k \partial^k (u^i \phi) \nonumber \\
&&+ \partial_k \partial^i (u^k \phi) - u^k u^\varepsilon (\partial_k \phi^2) \partial_\varepsilon (u^i \phi) - u^k u^\varepsilon \phi^2 \partial_k \partial_\varepsilon (u^i \phi)  \nonumber\\
&&- u^i u^\varepsilon (\partial_k \phi^2) \partial_\varepsilon (u^k \phi) - u^i u^\varepsilon \phi^2 \partial_k \partial_\varepsilon (u^k \phi), \nonumber \\
\\
y^i &=&  \big( \partial_i \partial_\varepsilon (u^{ \varepsilon} \phi) \big)  g^{i i} (1 - \phi^2 ) - \partial_\varepsilon (u^{ \varepsilon} \phi)  g^{i i} ( \partial_i \phi^2 ). \nonumber \\
\end{eqnarray*}

\noindent With the availability of Eqs.~(\ref{eq:conserve-energy}) and (\ref{eq:conserve-momentum}), 
the bulk and shear viscosity can be expressed explicitly, i.e.,
\begin{eqnarray}
\zeta = \frac{c^0 x^i - a^0 z^i}{b^0 x^i - a^0 y^i} ~.
\label{eq:bulk}
\end{eqnarray} 
\begin{eqnarray}
\eta = \frac{c^0 y^i - z^i b^0}{a^0 y^i - x^i b^0} ~. 
\label{eq:shear}
\end{eqnarray} 

The expression of the scalar field $\phi$ can be obtained by solving
Eq.~(\ref{eq:eq-of-motion}).  For simplicity, but without
losing generality,  the partial 
differential equation is assigned to live in two-dimension, $\phi=\phi(t,x)$, 
and the gauge bosons from each gauge group are assumed to be homogeneous. 
As a consequence, the differential equation becomes
\begin{equation}
\partial_t (\gamma \mathbf{v} \phi ) + \partial_x (\gamma \phi)   = -g_F  (J^a_{F 0} + F_0) ,
\label{eq:eq-of-motion2}
\end{equation}
with $\partial_t \equiv {\partial}/{\partial t}$, 
$\partial_{t x} \equiv {\partial^2}/{\partial t \partial x}$,  $g_F$ and $g_G$ 
are the gauge group SU(3) and U(1) coupling 
constants, respectively.
Here, $J^a_{F 0} = \bar{Q} \gamma_\mu T^a Q = Q^\dagger \gamma_0 \gamma_\mu T^a Q$, 
and 
\begin{eqnarray}
F_0 &=& i \gamma^2 \mathbf{v}^2 \phi \partial_t \phi - ({g_G}/{g_F}) A \gamma 
\partial_t \phi +  i \gamma^2 \mathbf{v}^2 \phi \partial_x \phi  - ({g_G}/{g_F}) A 
\gamma \partial_x \phi.
\end{eqnarray}
Thus, Eq.~(\ref{eq:eq-of-motion2}) can be written as
\begin{equation}
\delta \partial_{tx} \phi + \gamma \partial_{xx} \phi + \xi + \beta \phi 
\partial_t \phi + \beta \phi \partial_x \phi - \lambda \partial_t \phi - \lambda  
\partial_x \phi = 0,
\label{eq:parsial-dif}
\end{equation}
with $\delta = \gamma |\mathbf{v}|$, $\xi = g_F \bar{Q} \gamma_\mu T^a Q$ , 
$\beta = i \gamma^2 |\mathbf{v}|^2$, and $\lambda = (g_G / g_F) A \gamma$.
This equation can be solved analytically when it is converted to
an ordinary differential equation form 
by redefining  $\phi = \phi (x - iEt)$, 
where $ x - iEt = z$. Then, it follows that $\partial_x = \partial_z$, 
$\partial_t = - i E \partial_z$, $\partial_{tx} = -i E \partial_{zz}$, and 
$\partial_{xx} = \partial_{zz}$.
Therefore, Eq.~(\ref{eq:parsial-dif}) becomes,
\begin{eqnarray}
\alpha_1 \phi_{zz} - \alpha_2 \phi \phi_z + \alpha_3 \phi_z &=& \xi,
\end{eqnarray}
with $\alpha_1 \equiv - i E \delta + \gamma$, $\alpha_2 \equiv i E 
\beta - \beta$, and $\alpha_3 \equiv i E \lambda - \lambda$

Since the particular solution contributes only to transient condition, 
here we try to obtain an analytic homogeneous solution, 
\begin{equation}
\alpha_1 \phi_{zz} - \alpha_2 \phi \phi_z + \alpha_3 \phi_z = 0,
\end{equation}
and by defining 
\begin{equation*}
  \phi \equiv  \frac{\partial \omega}{\partial z} = \omega_z
\end{equation*}
we obtain
\begin{eqnarray}
\alpha_1 \omega_{zzz} - {\textstyle\frac{1}{2}}{\alpha_2} (\omega^2_z)_z  + 
\alpha_3 \omega_{zz} = 0 .
\label{eq:arranged}
\end{eqnarray}
Equation (\ref{eq:arranged}) can be integrated to obtain
\begin{eqnarray}
\int \frac{d \omega_z}{\left({\textstyle\frac{1}{2}}{\alpha_2} \omega^2_z - \alpha_3 
\omega_z\right)} = \frac{1}{\alpha_1} \int d z 
\end{eqnarray}
and by making use of the standard integral formula  
\begin{equation*}
  \int \frac{d x}{ b x^2 - a x}  
= \frac{ \ln [(b x - a)/x]}{a} ,
\end{equation*} 
we arrive at
\begin{eqnarray}
\phi = \omega_z = \frac{{2 \alpha_2 e^{C_1 \alpha_3}}}{{\alpha_2 
e^{C_1 \alpha_3} - 2 \; e^{{\alpha_3} z/{\alpha_1} + C_2 \alpha_3}}}.
\end{eqnarray}
For the sake of simplicity, 
the integration constants $C_1$ and $C_2$ are set to  
zero, which reduces the field to
\begin{equation}
\phi  = \frac{2 \alpha_2 }{{\alpha_2  - 2 e^{{\alpha_3} 
(x - i E t)/{\alpha_1}} }}.
\end{equation}

\section{Results and Discussion}
\label{sect:result}

\begin{figure}[!t]
\centerline{\includegraphics[width=8cm]{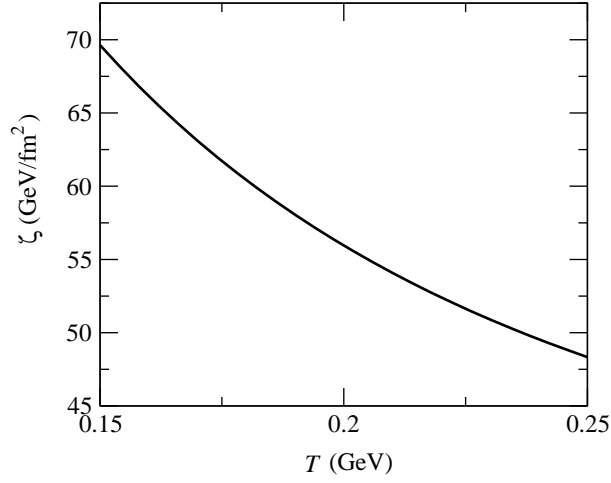}}
\caption{Bulk ($\zeta$) viscosity of 
      the gluon dominated QGP as a function of the temperature calculated
      from Eq.~(\ref{eq:bulk}). \label{fig:bulk}}
\end{figure}

\begin{figure}[t]
  \begin{center}
    \leavevmode
    \epsfig{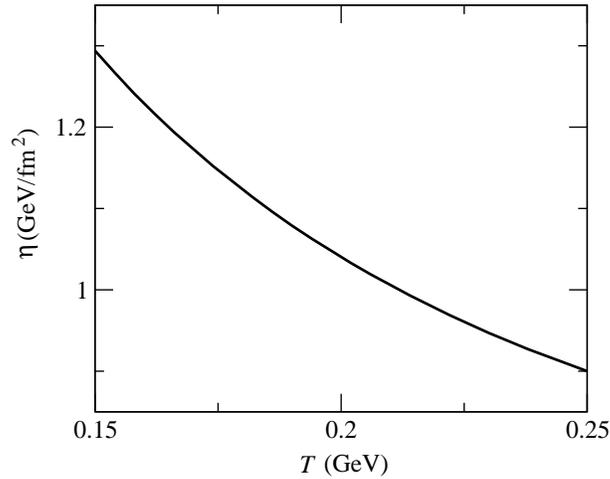}
    \caption{Shear ($\eta$) viscosity of 
      the gluon dominated QGP as a function of the temperature calculated
      from Eq.~(\ref{eq:shear}).}
   \label{fig:shear} 
  \end{center}
\end{figure}

To visualize our result for the gluon dominated QGP
at hadronization temperature, we take the temperature 
 range between 150 and 250 MeV. Note that 
150 MeV is the hadronization temperature that commonly adopted in the 
relativistic hydrodynamics description of QGP, while 
the values above 200 MeV are usually considered in the lattice QCD computation for 
QCD plasma. 
The other variable values are also chosen with respect to the condition 
where gluon dominated plasma is conjectured to exist. 
The particle velocity $v$ is 
assumed to be near the velocity of light, 
whereas $\Delta t = 1 \times 10^{-13}$ s.
The space variable $x$ is taken from  
$10^{-5}$ to $10^{-3}$ nm  and the quark mass $m_Q = 0.5$ GeV.
The energy of gauge field $A$ is set to $100$ GeV.
This value is taken with an assumption that the wave length of gauge field
is about two order smaller than the radius of quark $r_Q$,
where $r_Q \sim {\cal O} (10^{-15})$ m.
Then, by using 
$g_G \thicksim 1/137$ and $g_F \thicksim 1.00$, the ratio between 
weak coupling constant and strong coupling constant $g_G / g_F$ is 
found to be 0.0072. 
It should be noted that from all of the variable values where gluonic plasma
is conjectured to exist, the values of 
quark mass $m_Q$ and the energy of gauge bosons $A$ span a large range. 
However the result in the figures does not much depend
on the values of these variables. The reason is that the value
of $m_Q$ and $A$ can be reciprocally altered due to the assumption
that the higher the energy level of gauge boson 
in a circumstance, the smaller the possibility of heavy particles to exist.
If $A$ is set to a bigger value, 
then $m_Q$ has to be set to a smaller one, and vice versa.

Figure \ref{fig:bulk} shows that the  
obtained value for bulk viscosity increase, 
especially at $T$ very close to $T  \sim 0.15$ GeV. 
It is not totally similar with the result which is obtained from 
the calculation that utilizing quantum kinetic theory, 
where near to $T_c $ the bulk viscosity rises asymptotically. 

Figure \ref{fig:shear} exhibits the calculated shear viscosity, which 
has a small increment near the hadronization energy level. 
Compared to the bulk viscosity depicted in Fig.~\ref{fig:bulk}, 
the shear viscosity turns out to be one order smaller than 
the bulk viscosity.  
Qualitatively one can see that $\zeta \sim {\cal O} (10)$, and $\eta \sim {\cal O} (10^0)$.
With the conjectured value of shear viscosity to entropy ratio , $\eta / s \geq 1 / 4 \pi$,\cite{kovtun,kovtun1} 
(Kovtun-Son-Starinets bound in natural units)
one can conclude that the entropy of QGP is also 
of the order of $s \sim {\cal O} (10)$,
and $\zeta / s \gtrsim {\cal O} (10^0)$. 
This result is comparable to that of Tuchin {\it et. al.} for the 
QCD plasma, where $\zeta / s \sim 1$ \cite{tuchin,tuchin2},
which indicates that for a hot QCD matter 
there is a significant increase of the bulk viscosity at the 
vicinity of critical temperature, where in this case 
the adopted critical temperature $T_c$ is about $0.28$ GeV.  
It might be seen as an indication that at the hadronization process, 
bulk viscosity plays a more important role than shear viscosity. At least, 
mechanically, this large $\zeta$ brings the tendency that the fluidic QGP becomes 
unstable \cite{Bluhm:2010qf}. 

Comparison between Fig.~\ref{fig:bulk} and Fig.~\ref{fig:shear} 
reveals a linear relation between 
$\zeta$ and $\eta$ at about hadronization energy level.
It can be simply written as $\zeta \approx c \; \eta$, where $c \sim 50$. 
Equation (\ref{eq:energy-momentum-tensor_ideal}) shows that in the 
gluon fluidic system there exists no self-interaction between gluon fileds.
Instead, the interaction originates from the matter gluon relation.
The large value of $\zeta$ originates from the strong coupling constant 
in the gluon dominated system.
This is just the opposite to the huge ratio of $\eta / \zeta$  that 
appears at weakly coupled system \cite{dobado}.
This is also related to the fact that for any nonconformal field theory 
the bulk viscosity will always emerge and a non-Abelian gauge theory
like the gluon dominated QGP that is modeled in this discussion becomes 
nonconformal at hadronization energy level \cite{meyer2}.

\section{Summary}
\label{sect:summary}

Based on the first principle, an analysis started from the construction 
of a viscous fluid-like QCD Lagrangian has been performed. 
The phase transition between stable hadronic state and highly 
energized QGP state has been investigated. It is argued that 
inside the hadronic state the gluons behave as point particles 
and their properties are determined by their polarization vectors. 
However, in the hot QGP state the gluons should behave as fluid 
particles and characterized by their relativistic velocities and their 
inherently attached scalar fields. By investigating the energy-momentum 
tensor for viscous fluid particles,  a detailed derivation of the bulk 
and shear viscosities  close to the deconfinement temperature 
has been performed. The result of this investigation corroborates 
the findings obtained in the previous studies that make use of 
the QCD  low-energy-theorem approach. 

Another interesting issue related to the result of the present analysis  
is the calculation of cross-section. In our approach presented in this paper, 
which is based on hydrodynamics, calculation of the cross-section seems to be
difficult.
In principle, it would be possible to relate the result obtained 
for the viscosity with the prediction of kinetic theories, where the transport
coefficients like bulk and shear viscosities in a gluon dominated QGP
are connected to the gluon cross-section \cite{plumari}. 
Nevertheless, this would require a much more 
comprehensive and careful investigation,
which is planed for the future works.

\section{Acknowledgment}
T.P.D. thanks A. Sulaiman and H. S. Ramadhan for fruitful discussion 
and the Group for Theoretical and Computational Physics, LIPI, 
for the warm hospitality extended to him. T.P.D. and L.T.H. are 
supported by Riset Kompetitif LIPI under Contract 
No. 11.04/SK/KPPI/II/2014. T.M. is supported by the 
Research-Cluster-Grant-Program of the University of Indonesia, 
contract No. 1709/H2.R12/HKP.05.00/2014.



\begin{thebibliography}{0}    
\bibitem{Gottlieb:2007zz} 
  S.~Gottlieb, 
  J.\ Phys.\ Conf.\ Ser.\  {\bf 78}, 012023 (2007).
\bibitem{Petreczky:2007bn} 
  P.~Petreczky, 
  Eur.\ Phys.\ J.\ ST {\bf 155}, 123 (2008).
\bibitem{Bouras:2009nn} 
  I.~Bouras, E.~Molnar, H.~Niemi, Z.~Xu, A.~El, O.~Fochler, C.~Greiner 
  and D.~H.~Rischke, 
  Phys.\ Rev.\ Lett.\  {\bf 103}, 032301 (2009).
\bibitem{Romatschke:2009im} 
  P.~Romatschke,
  Int.\ J.\ Mod.\ Phys.\ E {\bf 19}, 1 (2010).
\bibitem{Teaney:2000cw} 
  D.~Teaney, J.~Lauret and E.~V.~Shuryak,
  Phys.\ Rev.\ Lett.\  {\bf 86}, 4783 (2001).
\bibitem{Huovinen:2001cy} 
  P.~Huovinen, P.~F.~Kolb, U.~W.~Heinz, P.~V.~Ruuskanen and S.~A.~Voloshin,
  Phys.\ Lett.\ B {\bf 503}, 58 (2001).
\bibitem{Kolb:2001qz} 
  P.~F.~Kolb, U.~W.~Heinz, P.~Huovinen, K.~J.~Eskola and K.~Tuominen,
  Nucl.\ Phys.\ A {\bf 696}, 197 (2001).
\bibitem{Kolb:2002ve} 
  P.~F.~Kolb and R.~Rapp,
  Phys.\ Rev.\ C {\bf 67}, 044903 (2003).
\bibitem{Hirano:2002ds} 
  T.~Hirano and K.~Tsuda,
  Phys.\ Rev.\ C {\bf 66}, 054905 (2002).
\bibitem{Baier:2006gy} 
  R.~Baier and P.~Romatschke,
  Eur.\ Phys.\ J.\ C {\bf 51}, 677 (2007).
\bibitem{Bluhm:2010qf} 
  M.~Bluhm, B.~K\"ampfer and K.~Redlich,
  Phys.\ Rev.\ C {\bf 84}, 025201 (2011).
\bibitem{Adams:2003kv} 
  J.~Adams {\it et al.}  
  Phys.\ Rev.\ Lett.\  {\bf 91}, 172302 (2003).
\bibitem{Adare:2008qa} 
  A.~Adare {\it et al.}  
  Phys.\ Rev.\ Lett.\  {\bf 101}, 232301 (2008).
\bibitem{Bambah:2006yg} 
  B.~A.~Bambah, S.~M.~Mahajan and C.~Mukku,
  Phys.\ Rev.\ Lett.\  {\bf 97}, 072301 (2006).
  
\bibitem{jackiw}
  R.~Jackiw, S.~Y.~Pi and A.~P.~Polychronakos,
  Annals Phys.\  {\bf 301}, 157 (2002).

\bibitem{jackiw2}
  R.~Jackiw, V.~P.~Nair, S.~Y.~Pi and A.~P.~Polychronakos,
  J.\ Phys.\ A {\bf 37}, R327 (2004).

\bibitem{sulaiman} 
  A. Sulaiman, A. Fajarudin, T.P. Djun and L.T. Handoko,
 Int. J. Mod. Phys. A {\bf 24}, 3630 (2009).

\bibitem{djun} 
  T.P. Djun and L.T. Handoko,
  {\it Proceeding of the Conference in Honour of Murray Gell-Mann's 80th Birthday: 
  Quantum Mechanics, Elementary Particles, Quantum Cosmology and Complexity;
  ``Fluid QCD approach for quark-gluon plasma in stellar structure,''}
  arXiv:1109.6066 [hep-ph].

\bibitem{landau} L. D. Landau and E. M. Lifshitz, {\it Fluid Mechanics} 
  (Butterworth-Heinemann, Oxford, 2003) 2nd Ed.

\bibitem{weinberg} S. Weinberg, {\it Gravitation and Cosmology}
  (John Wiley \& Sons, New York, 1972).  

\bibitem{nugroho} 
  C.S. Nugroho, A.O. Latief, T.P. Djun and L.T. Handoko,
  Grav. Cosmol. {\bf 18}, 32 (2012).

\bibitem{ryder} L. Ryder, {\it Introduction to General Relativity}
  (Cambridge University Press, Cambridge, 2009). 




\bibitem{tuchin}
  D.~Kharzeev and K.~Tuchin,
  JHEP {\bf 0809}, 093 (2008).
  
\bibitem{tuchin2}
  F.~Karsch, D.~Kharzeev and K.~Tuchin,
  Phys.\ Lett.\ B {\bf 663}, 217 (2008).

\bibitem{kovtun} 
  P. Kovtun, D.T. Son  and A.O. Starinets,
  Phys. Rev. Lett.  {\bf 94}, 111601 (2005).

\bibitem{kovtun1} 
  P. Kovtun, D.T. Son  and A.O. Starinets,
  JHEP {\bf 0310},  064 (2003).
  
\bibitem{dobado}
  A. Dobado, F.J. Llanes-Estrada and J.M. Torres-Rincon
  Phys.\ Rev.\ D {\bf 88}, 0539006 (2013).
  
\bibitem{meyer2} 
  H.B. Meyer,
  {\it Phys. Rev. Lett.}  {\bf 100}, 162001 (2008).
  
\bibitem{plumari} 
  S.~Plumari, A.~Puglisi, F.~Scardina and V.~Greco,
  Phys.\ Rev.\ C {\bf 86}, 054902 (2012).  
 
\end{thebibliography}
\end{document}